\documentclass[10pt]{article}
\pdfoutput=1
\usepackage[dvips]{graphicx}
\usepackage{amsthm}
\usepackage{epsfig} 
\usepackage{amssymb}
\usepackage{url}
\usepackage{color}

\def\be{\begin{equation}}
\def\bea{\begin{eqnarray}}
\def\ee{\end{equation}}
\def\eea{\end{eqnarray}}

\def\apx{\approx}

\newenvironment{arxivabstract}{%
\begin{quote} \bf}
{\end{quote}}

\newcounter{lastnote}

\title{Limit Sets for Natural Extensions of Schelling's Segregation Model}

\author
{Abhinav Singh,$^{1}$ Dmitri Vainchtein,$^{2,}$$^{3\ast}$ Howard Weiss$^{4}$\\
\\
\normalsize{$^{1}$School of Physics and Center for Nonlinear Science, Georgia Tech, USA}\\
\normalsize{$^{2}$Department of Mechanical Engineering, Temple University, USA}\\
\normalsize{$^{3}$Space Research Institute, Moscow, Russia}\\
\normalsize{$^{4}$School of Mathematics and Center for Nonlinear Science, Georgia Tech, USA}\\
\\
\normalsize{$^\ast$To whom correspondence should be addressed;
E-mail: dmitri@temple.edu.}}

\date{}

\begin{document}

\maketitle

\begin{arxivabstract}

Thomas Schelling developed an influential demographic model that illustrated how, even with relatively mild assumptions on each individual's nearest neighbor preferences, an integrated city would likely unravel to a segregated city, even if all individuals prefer integration. Individuals in Schelling's model cities are divided into two groups of equal number and each individual is 'happy' or 'unhappy' when the number of similar neighbors cross a simple threshold. In this manuscript we consider natural extensions of Schelling's original model to allow the two groups have different sizes and to allow different notions of happiness of an individual. We observe that differences in aggregation patterns of majority and minority groups are highly sensitive to the happiness threshold; for low threshold, the differences are small, and when the threshold is raised, striking new patterns emerge. We also observe that when individuals strongly prefer to live integrated neighborhoods, the final states exhibit a new tessellated-like structure.

\end{arxivabstract}


\section{Introduction}

In the 1970s, the eminent economic modeler Thomas Schelling proposed a simple space-time population model to illustrate how, even with relatively mild assumptions concerning every individual's nearest neighbor preferences, an integrated city would likely unravel to a segregated city, even if all individuals prefer integration \cite{S1,S2,S3,S4}. Individuals in Schelling's cities are divided into two groups of equal number and each individual is 'happy' or 'unhappy' when the number of similar neighbors cross a simple threshold. This agent based lattice model has become quite influential amongst social scientists, demographers, and economists, and some authors have used the Schelling-like models to analyze actual populations in cities \cite{Clark,BR,BOHO,SSD,ClarkFossett}. Currently, there is a spirited discussion amongst demographers on the validity of Schelling-type models to describe actual segregation, with arguments both for (e.g., \cite{Young98, Fossett06}), and against (e.g., \cite{MasseyDenton93, LaurieJaggi}).

Aggregation relates to individuals from the same group joining together to form clusters. Schelling equated global aggregation with segregation. Many authors assumed that the striking global aggregation observed in simulations on very small ideal ``cities" persists for large, realistic size cities. In \cite{SVW09} we showed that this is false. There have been simulations of segregation models for large cities, in part due to the large computational costs required to run simulations using existing algorithms \cite{PW, Fossett06, Fossett06, Zhang04, VK}. We developed highly efficient and fast algorithms that allow us to run many simulations for many sets of parameters and to compute meaningful statistics of the measures of aggregation.

We modify two central assumptions of Schelling's original model. Schelling assumed that the number of agents in both groups is the same and we allow different numbers (a majority and a minority). Schelling also defined an agent as being either 'happy' or 'unhappy' based on a threshold number of agents from the same group in its neighborhood, and we consider two new happiness criteria: 1) the happiness of an agent is a linearly increasing function of the number of similar agents in its neighborhood, and 2) an agent is maximally happy in a completely integrated neighborhood and its happiness declines linearly when the neighborhood is dominated by either type of agents (see Figure~\ref{utility}).

We show that the happiness threshold plays an important role in cities where one group forms a majority. When an agent needs three similar agents in its neighborhood to be happy, there is little difference in the aggregation patterns of majority and minority agents. When the threshold rises to four, distinct geometric differences emerge. When agents prefer to live in integrated neighborhoods, the two types of agents arrange themselves in a tessellated-like structure across the city.



{\color{red} }

\subsection{Description of the Model}




We follow \cite{SVW09} and view Schelling's model \footnote{Different authors frequently consider slightly different versions of Schelling's original model, i.e.,different ways of moving boundary agents. All versions seem to exhibit the same qualitative behaviors, and thus we refer to {\it the} Schelling model.} as a three parameter family of models. The phase space for these models is the $N\times N$ square lattice with periodic boundary conditions (opposite sides identified). We consider two distinct populations composed of black agents (squares) $B$ and red agents (squares) $R$ (red squares appear grey on b/w printing) and we do not assume that $\# B=\# R$. Together these agents fill up most of the $N^2$ sites, with $V$ remaining vacant sites (white squares). Each agent has eight nearest neighbors, corresponding to a Moore, or Queen, neighborhood. Different types of neighborhoods were considered by different authors (see, e.g, \cite{Fossett06,ClarkFossett}, where the size of the neighborhood was referred to as `vision'). Demographically, the parameter $N$ controls the size of the city and $v=V/N^2$ controls the population density or the {\it occupancy ratio} \cite{realestate}. We introduce a utility function, $U_{i,j}$ that measures the {\it happiness} of the agent at lattice square $({i,j}) $ as a function of the states of its eight nearest neighbors. The function $U$ can have two (i.e., $0$ and $1$ -- "unhappy" and "happy") or more values. The convention is that larger values of $U$ for a given agent correspond to increased happiness.

We follow Schelling and begin the evolution by choosing an initial configuration starting with a checkerboard with periodic boundary conditions. Then, if necessary, we substitute $B$ agents by $R$ agents to achieve the desired ratio $\# R/ \# B$. Demographically, a checkerboard configuration is a maximally integrated configuration. We then randomly remove $\#V$ agents to create vacant locations (keeping the ratio $\# R/ \# B$ constant). Finally we permute agents in two $3 \times 3$ blocks. Alternatively, we could choose a random initial configuration. In general, except for small values of $v$, the final states with a random initial configuration are quantitatively similar to the ones obtained using the Schelling-like initial conditions.

We randomly select a $B$ agent and a vacant site, such that when moved to the vacant site the $B$ agent becomes ``happier''. If the utility function $U$ only attains the values $0$ and $1$, corresponding to "unhappy" and "happy" as in the original Schelling protocol (\cite{S1,BP94,BOHO,SVW09}), the $B$ agent must be unhappy at the original location and happy at the new location. Provided this is possible, we interchange the $B$ agent with the vacant site, so that the utility function of the $B$ agent increases. Then we randomly select an $R$ agent and a vacant site, where that $R$ agent would be happier by switching with the vacant site. Provided this is possible, we interchange the $R$ with the vacant site. We repeat this iterative procedure, alternating between selecting a $B$ agent and an $R$ agent, until a {\it final state} is reached, where no interchange is possible that increases happiness. For some final states, some (and in some cases, many) agents may be unhappy, but there are no allowable switches.

We simulate the model and quantify the aggregation. We currently need approximately one minute to run a single simulation for a city of size $N=100$ and we ran thousands of simulations for this manuscript. The details of the algorithm were presented in \cite{SVW09}. We study the dynamics for large lattices and present our results for $\mbox{city size}~N=100$. As in \cite{SVW09}, choosing $N$ greater than $100$ does not lead to qualitatively or quantitatively different states and phenomena.

\section{Minorities}

We first consider an extension of the original Schelling model to allow for "minority" and majority populations -- configurations where the number of $R$ agents is larger that the number of $B$ agents, or visa versa. The ``agent comfortability index'', $T\in \{0, 1, \dots, 8\}$, quantifies an agent's tolerance to living amongst disparate nearest neighbors. For a given value of $T$, a $B$ or $R$ agent is {\it happy} if $T$ or more of its nearest eight neighbors are $B$'s or $R$'s, respectively. Else it is unhappy. We follow Schelling's evolution algorithm \cite{S1}, later used in \cite{BP94,BOHO}, and begin by choosing an initial configuration by the method described above. We then randomly select an unhappy $B$ and a vacant site surrounded by at least $T$ nearest $B$ neighbors. Provided this is possible, we interchange the unhappy $B$ with the vacant site, so that this $B$ becomes happy. We then randomly select an unhappy $R$ and a vacant site having at least $T$ nearest neighbors of type $R$. Provided this is possible, we interchange the unhappy $R$ and the vacant site, so that $R$ becomes happy. We repeat this procedure, alternating between selecting an unhappy $B$ and an unhappy $R$, until a {\it final state} is reached, where no interchange is possible that increases happiness. For some final states, some (and in some cases, many) agents may be unhappy, but there are no allowable switches.

To quantify the disparity between the number of agents, we introduce the parameter
\[
r=\frac{\# R}{\#R + \# B}.
\]
Without loss of generality we assume that $\#R \ge \# B$, so that, $0.5 \le r \le 1$. The case $r=0.5$ corresponds to the equal numbers of agents and $r=1$ corresponds to all red agents. Numerical simulations indicate that meaningful results only occur for $r$ values between $0.5$ and $0.7$. For larger $r$ values the minority agents are too far apart and can not provide sufficient nuclei for aggregation.

We consider $\mbox{neighbor comfort thresholds}~T=3, 4$ and $\mbox{vacancy ratio}~v$ between $2\%$ and $33\%$. The system does not evolve very much for other values of $T$: for $T=1, 2$ almost all of the agents are satisfied in most of the initial configurations, while for $T \ge 5$ there are almost no legal switches for the minority agents. Values of $v$ larger than $33\%$ correspond to unrealistic environments. For each pair of parameters $T$ and $v$, we perform $100$ simulations and we determine mean values of aggregation measures based on these 100 simulations. As our sample size (100) is large, the Central Limit Theorem provides 95\% confidence intervals for our estimates of aggregation measures.

Similarly to our construction in the $r=1/2$ case, we introduce {\it the adjusted perimeter per agent $p$ of the interface between the different agents suitably adjusted for the vacant spaces.} The perimeter $P$ is defined as twice the total number of $R$-$B$ connections plus the total number of connections between $R$ and $B$ agents with vacant spaces. Demographically, the adjusted perimeter, $p=P/N^2$, is the average number of contacts an agent has with the opposite kind or with vacant sites. In the segregation literature, the perimeter is related to the {\it exposure index} (see, e.g., \cite{MasseyDenton88}).

Our key observation is that $p$ is a {\it Lyapunov function}, i.e., a function defined on every configuration that is strictly decreasing along the evolution of the system. Thus the system evolves to minimize the adjusted interface between the $R$ and $B$ agents. The final states are precisely the local minimizers of the Lyapunov function, subject to the threshold constraint. This Lyapunov function is also the Hamiltonian for a related spin lattice system related to the Ising model \cite{Simon}. Such a definition of $p$ was motivated by analogies of these models with the physics of foams. Note, that for the triangular utility function, like the ones considered in \cite{Zhang04} and Sect.~3.2 below, $p$ is not a Lyapunov function.

In Figs.~\ref{T3r5}-\ref{T4r7}, we present characteristic final states for different values of $T$, $r$, and $v$. For the sake of comparison we include the corresponding figures for $r=0.5$ from \cite{SVW09}.


\subsection{$\mathbf{T=3}$}

\begin{figure}[htbp]
  \begin{center}
    \includegraphics[width=5in]{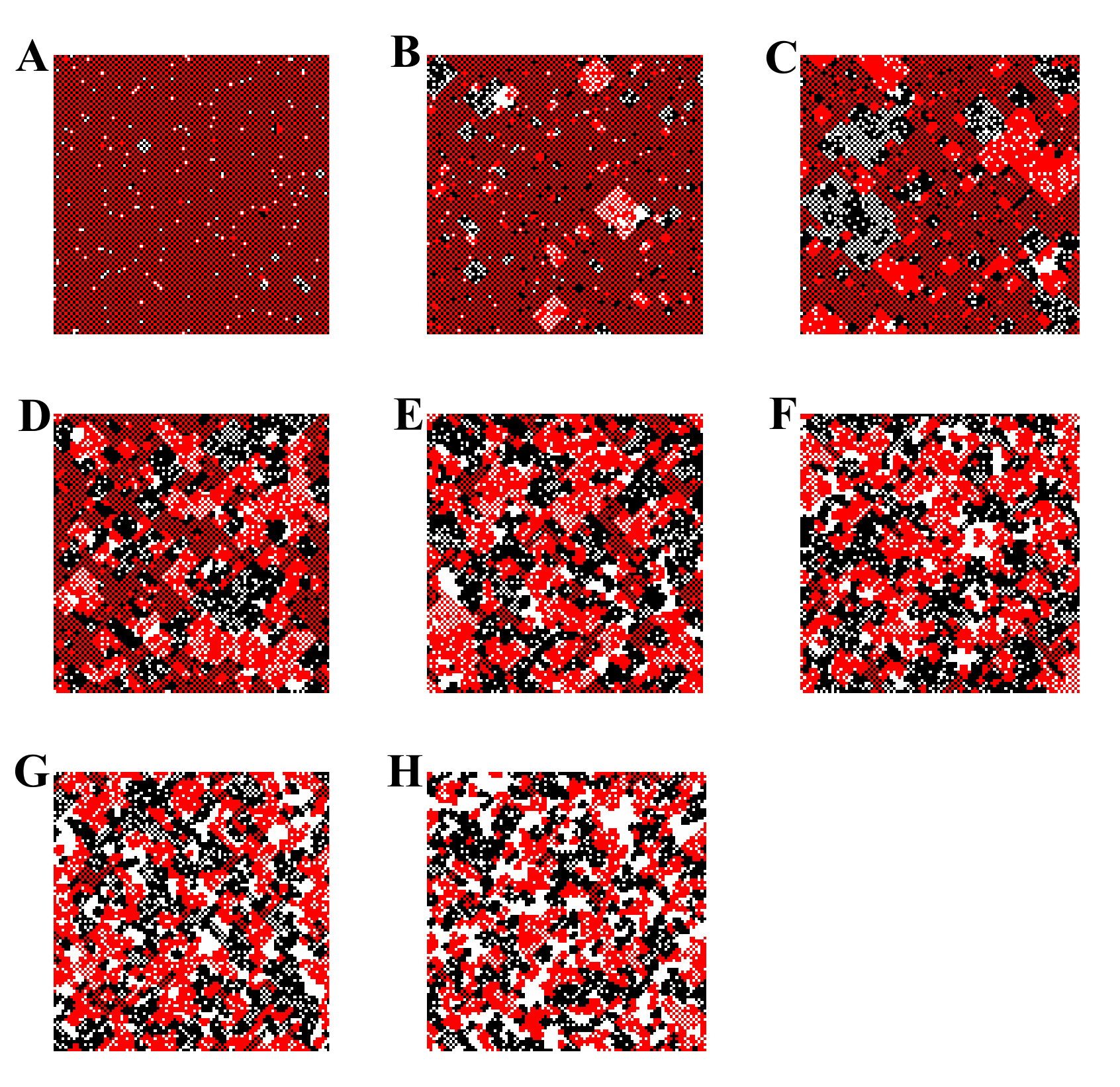}
\caption{\label{T3r5} Characteristic final states for neighbor comfort threshold $T=3$ and $r=0.5$ for
different vacancy ratio $v$: {\bf A}: $v=2\%$, {\bf B}: $v=6\%$, {\bf C}: $v=10\%$, {\bf D}:
$v=15\%$, {\bf E}: $v=20\%$, {\bf F}: $v=24\%$, {\bf G}: $v=28\%$, {\bf H}:
$v=33\%$.}
  \end{center}
\end{figure}

\begin{figure}[htbp]
  \begin{center}
    \includegraphics[width=5in]{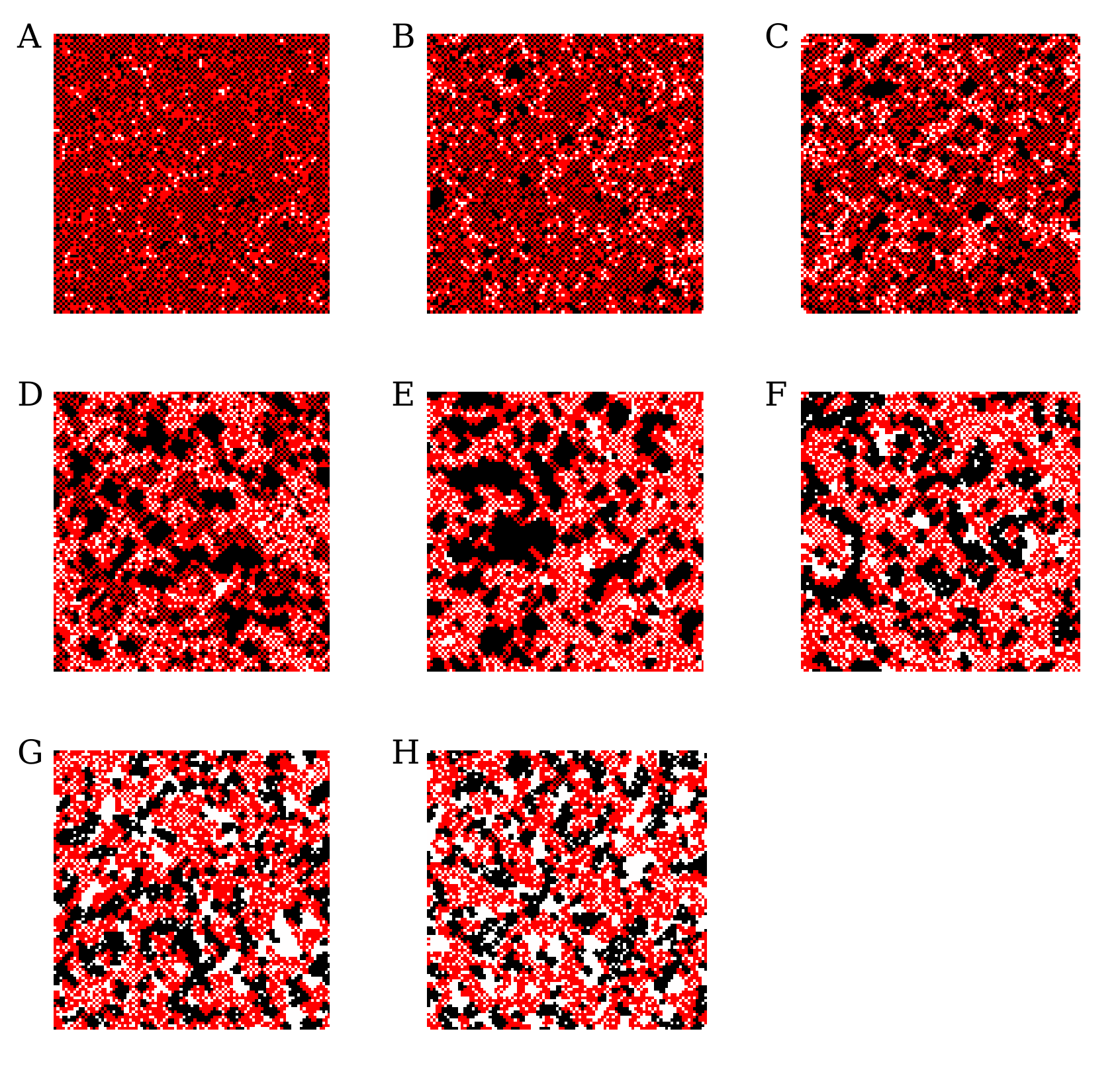}
\caption{\label{T3r6} Characteristic final states for neighbor comfort threshold $T=3$ and $r=0.6$ for
different vacancy ratio $v$: {\bf A}: $v=2\%$, {\bf B}: $v=6\%$, {\bf C}: $v=10\%$, {\bf D}:
$v=15\%$, {\bf E}: $v=20\%$, {\bf F}: $v=24\%$, {\bf G}: $v=28\%$, {\bf H}:
$v=33\%$.}
  \end{center}
\end{figure}

\begin{figure}[htbp]
  \begin{center}
    \includegraphics[width=5in]{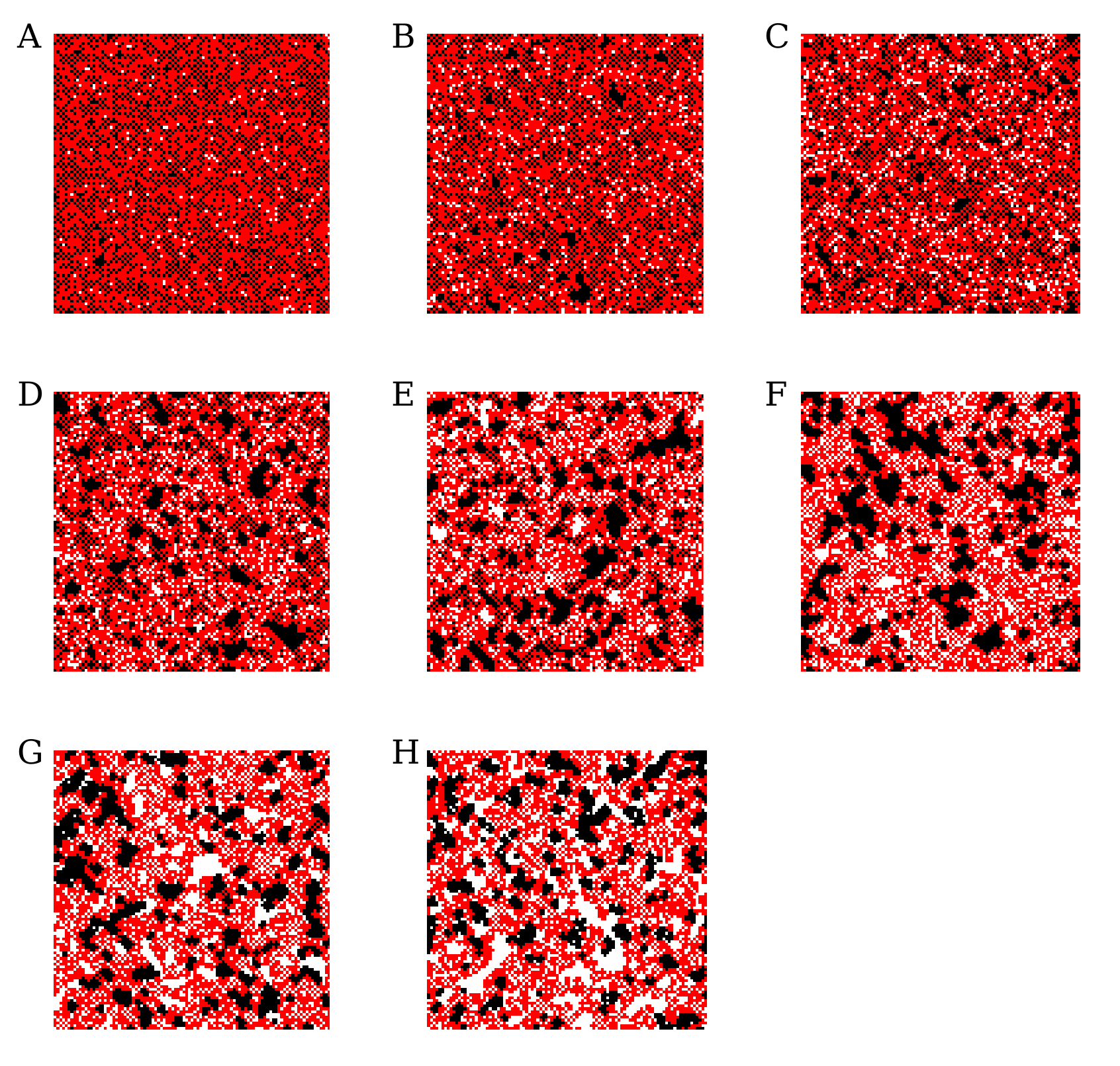}
\caption{\label{T3r7} Characteristic final states for neighbor comfort threshold $T=3$ and $r=0.7$ for
different vacancy ratio $v$: {\bf A}: $v=2\%$, {\bf B}: $v=6\%$, {\bf C}: $v=10\%$, {\bf D}:
$v=15\%$, {\bf E}: $v=20\%$, {\bf F}: $v=24\%$, {\bf G}: $v=28\%$, {\bf H}:
$v=33\%$.}
  \end{center}
\end{figure}

Figs.~\ref{T3r5}-\ref{T3r7} show characteristic final states for different values of $T=3$.

For small values of $v$ large blocks of the initial checkerboard configuration remain unchanged during the evolution. In \cite{SVW09} we called this phenomenon {\it super-stability} of the checkerboard. Every agent in a checkerboard is not just happy, it has four like neighbors; therefore has one like neighbor to spare. Thus it takes a large deviation from the checkerboard pattern to make an agent move and only agents close to the initially perturbed sites move. For the Minority agents the super-stability is less pronounced: as the Minority agents occupy way less than half of the squares, some of them in an original configuration have $3$, or even $2$, like neighbors. Therefore, the Minority agents are more sensitive to the perturbations of initial structure. This results in the appearance of small dense clusters of minority agents. The number of such clusters is smaller for $r=0.7$ than for $r=0.6$ because in the former case there are less $B$ agents. Otherwise the minority states do not differ much from the $r=0.5$ states.

\subsection{$\mathbf{T=4}$}
\begin{figure}[htbp]
  \begin{center}
    \includegraphics[width=5in]{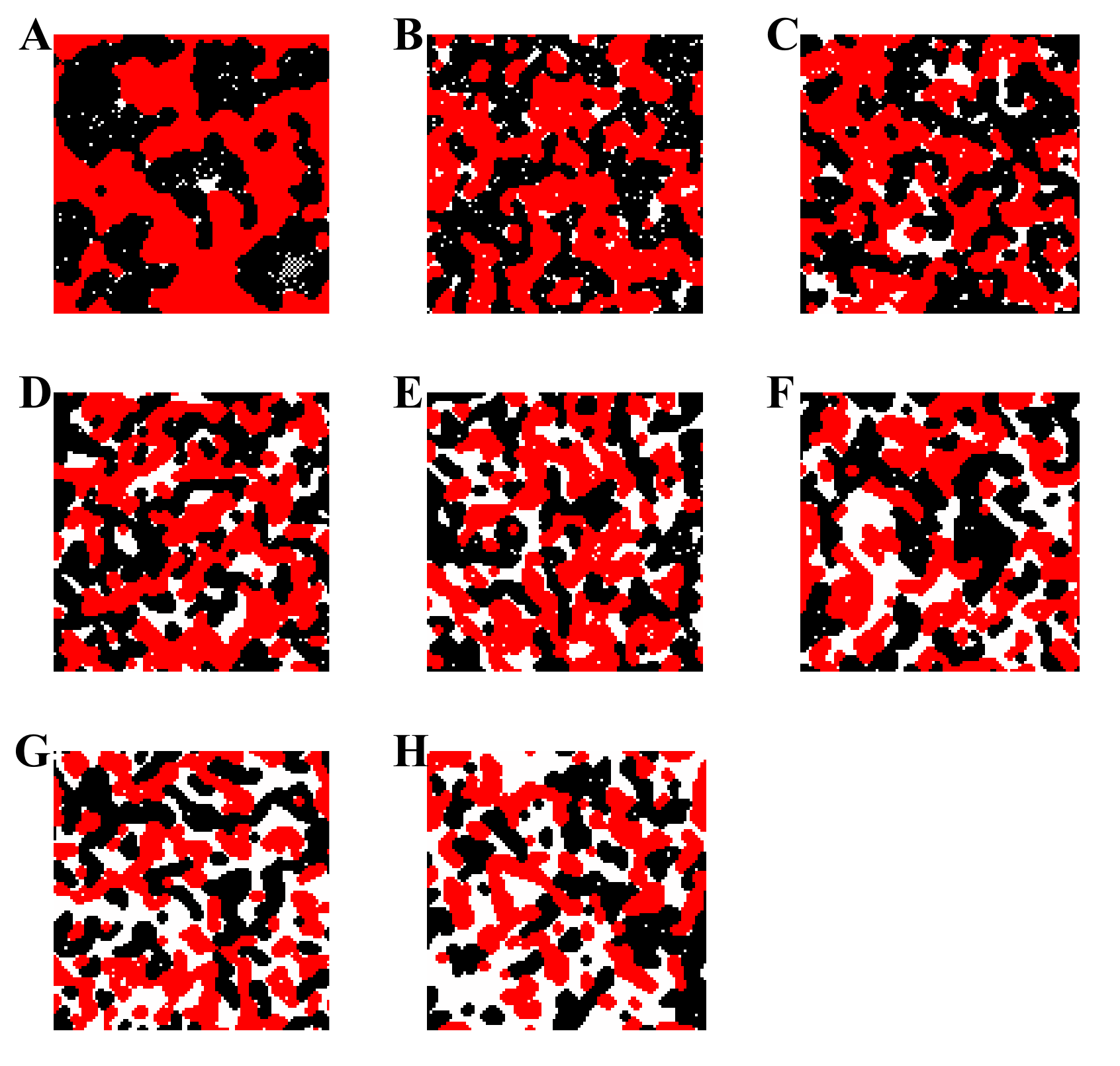}
\caption{\label{T4r5} Characteristic final states for neighbor comfort threshold $T=4$ and $r=0.5$ for
different vacancy ratio $v$: {\bf A}: $v=2\%$, {\bf B}: $v=6\%$, {\bf C}: $v=10\%$, {\bf D}:
$v=15\%$, {\bf E}: $v=20\%$, {\bf F}: $v=24\%$, {\bf G}: $v=28\%$, {\bf H}:
$v=33\%$.}
  \end{center}
\end{figure}

\begin{figure}[htbp]
  \begin{center}
    \includegraphics[width=5in]{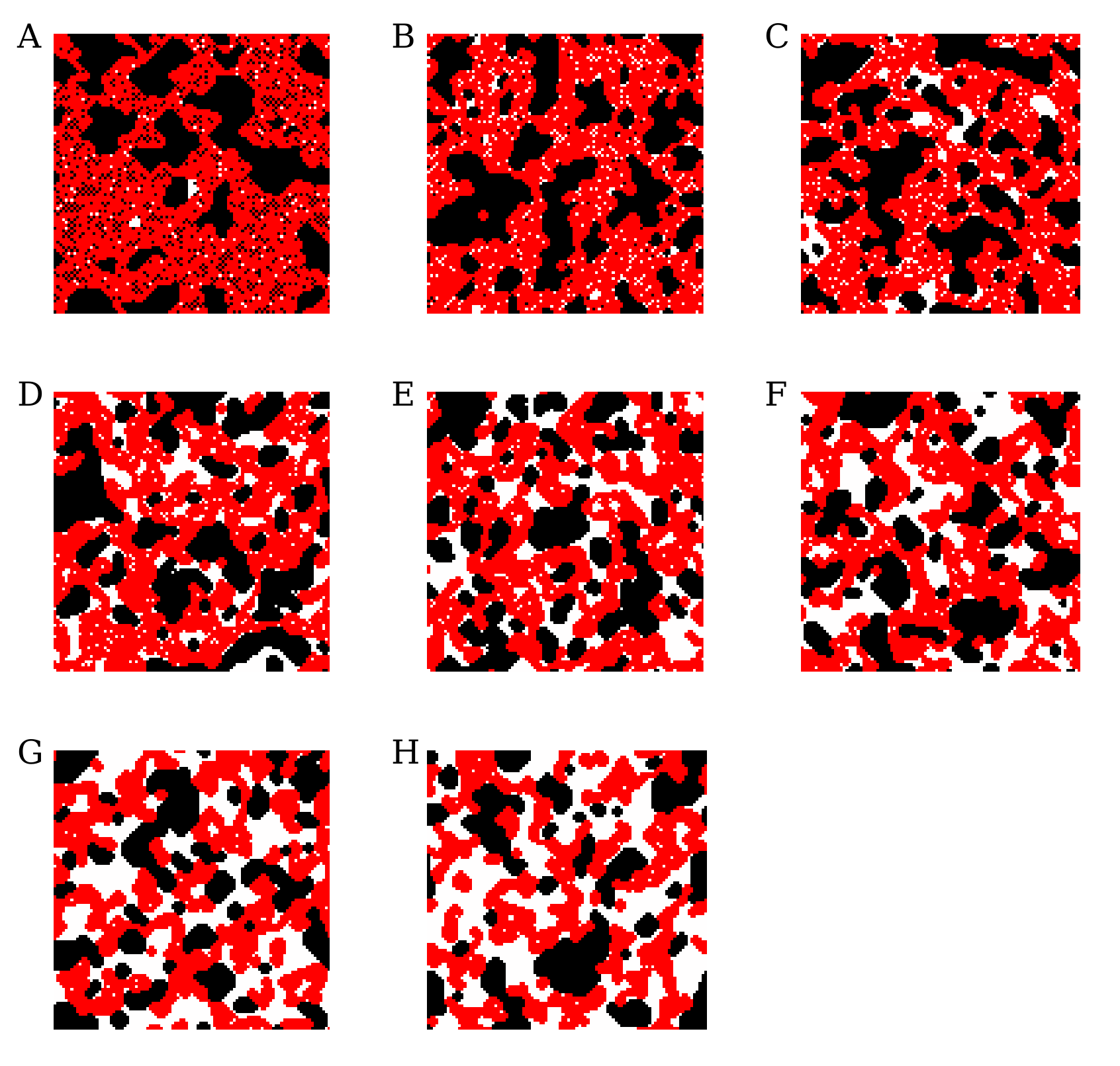}
\caption{\label{T4r6} Characteristic final states for neighbor comfort threshold $T=4$ and $r=0.6$ for
different vacancy ratio $v$: {\bf A}: $v=2\%$, {\bf B}: $v=6\%$, {\bf C}: $v=10\%$, {\bf D}:
$v=15\%$, {\bf E}: $v=20\%$, {\bf F}: $v=24\%$, {\bf G}: $v=28\%$, {\bf H}:
$v=33\%$.}
  \end{center}
\end{figure}

\begin{figure}[htbp]
  \begin{center}
    \includegraphics[width=5in]{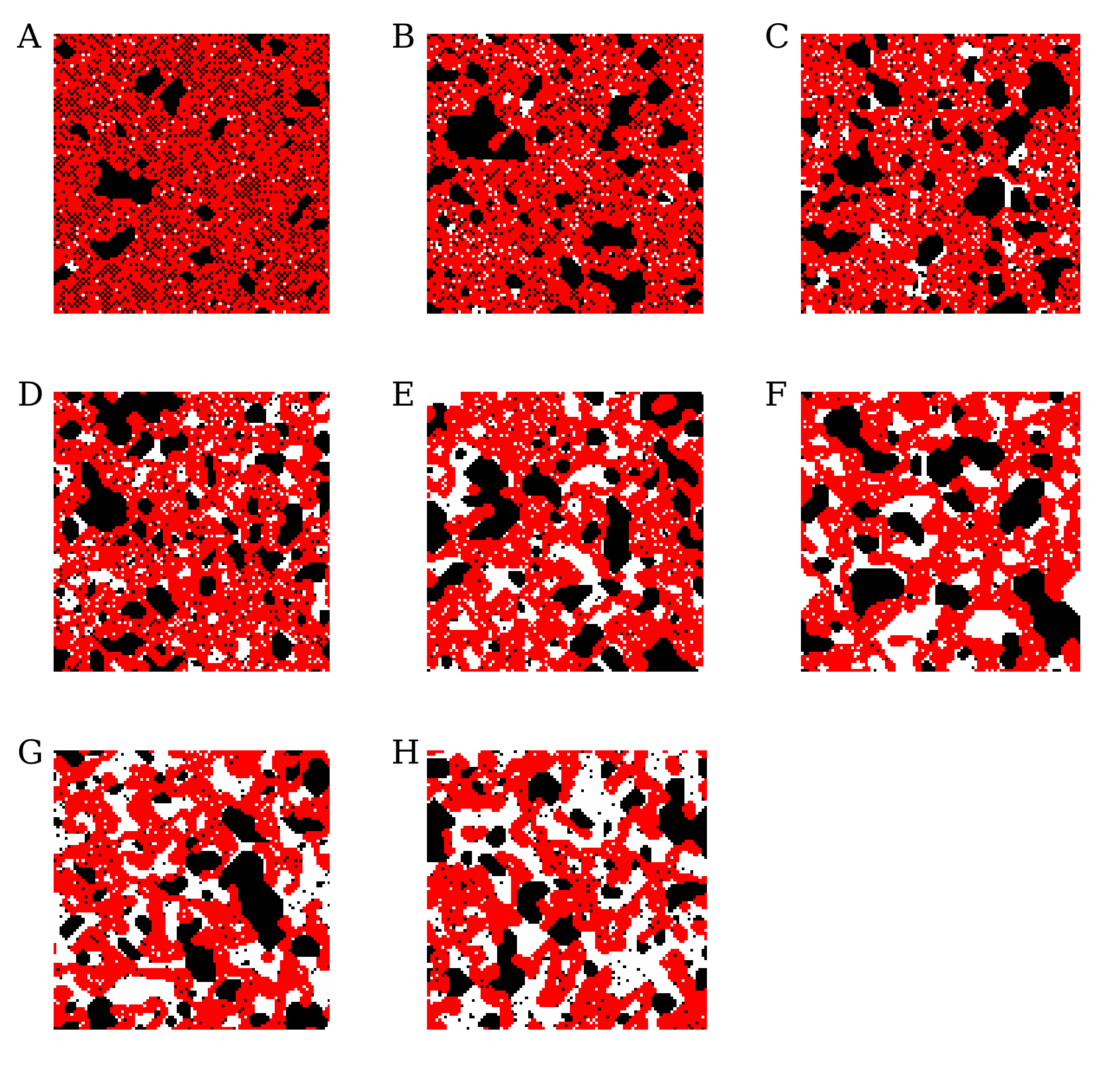}
\caption{\label{T4r7} Characteristic final states for neighbor comfort threshold $T=4$ and $r=0.7$ for
different vacancy ratio $v$: {\bf A}: $v=2\%$, {\bf B}: $v=6\%$, {\bf C}: $v=10\%$, {\bf D}:
$v=15\%$, {\bf E}: $v=20\%$, {\bf F}: $v=24\%$, {\bf G}: $v=28\%$, {\bf H}:
$v=33\%$.}
  \end{center}
\end{figure}

Figs.~\ref{T4r5}-\ref{T4r7} show characteristic final states for different values of $T=4$.

Unlike the case $r=0.5$, for larger values of $r$ there are unhappy minority agents in the final configurations. For $r=0.6$ the unhappy agents are present for $v=2\%$ only. For $r=0.7$ they are present all the way up to $v=33\%$, but their number steadily decreases as $v$ increases.

The major difference between the $T=4$ and $T=3$ cases is that for $T=4$, just by looking at the final state one can readily say which type of agents are in minority. For small values of $v$, the majority agents appear to be uniformly distributed over the city, while the minority agents are concentrated in a relatively few dense clusters. This phenomenon can be explained by the fact that in the initial configuration, even for small values of $v$, many minority agents are unhappy.

Similarly to the $T=3$ case, the distribution of the majority, $R$, agents remains almost the same as in the equal number case. They form dense clusters (almost no vacancies inside clusters) and the clusters are "snakelike": long and wavy, with a relatively large boundary to area ratio. The $B$ agents form smaller clusters, that are more "circular". These clusters also also uniformly distributed over the city.

\section{Alternative Utility Functions}

In our previous manuscript \cite{SVW09} we studied the final states in the Schelling Model with fixed threshold $T$. Here we study dynamics using linear and triangular utility functions. Thus the happiness of an agent is no longer a binary function. For the former, agents move as long as their happiness increases.

\begin{figure}[htbp]
  \begin{center}
    \includegraphics[width=3in]{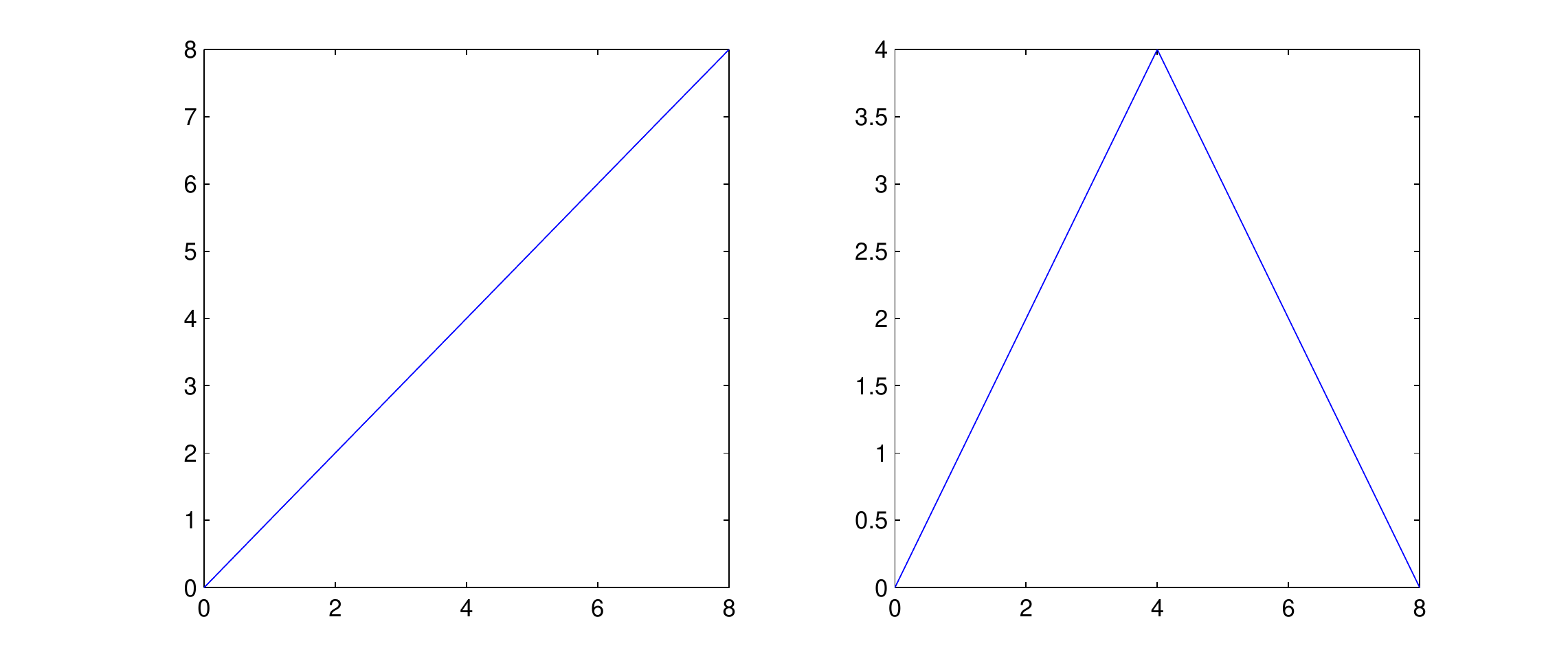}
\caption{\label{utility} Militancy (left panel) and Triangle (right panel) Utility functions.}
  \end{center}
\end{figure}

The linear utility function
\[
U_M = \#(\mbox{like neighbors}),
\]
corresponds to the desire of agents to be surrounded by as many similar agents as possible. The triangle utility function is in a sense opposite --
\[
U_T = 4 - \left| 4- \#(\mbox{like neighbors}) \right|,
\]
where the happiness increases linearly until an agent has four similar neighbors and then the happiness declines linearly to $0$. This is a particular case of mixed-neighborhood preferences (see, e.g., \cite{Sullivan:2009}. Thus an agent is maximally happy when it is surrounded by four similar neighbors. Such agents prefer to live in maximally mixed neighborhoods. The plots of the two utility functions are presented in Fig.~\ref{utility}.

We quantify the aggregation in final states using the four quantitative measures that we used in \cite{SVW09}:

\noindent (1) The $[u/l]$-measure is the ratio of unlike to like neighbors. For a lattice site with coordinates $(i, j)$ we define:
\[
[u/l]_{i,j} = \frac{q_{i,j}+w_{i,j}}{s_{i,j}},
\]
where $s_{i,j}$, $q_{i,j}$, and $w_{i,j}$ are the number of like, unlike, and vacant neighbors of the agent located at $(i,j)$, respectively. We define the {\it sparsity} $\left<[u/l]\right>$ of a cluster by averaging the $[u/l]$-measure over the given cluster.

\medskip

\noindent (2) {\it The number of agents that have neighbors {\it only} of the same kind} (note, that this definition excludes the vacant spaces). The abundance of such agents indicates the presence of large, ``solid'' clusters. This quantity is the most useful in distinguishing between the states with $T=3$ and $T=4$. We call the latter quantity the {\it seclusiveness}.

\medskip

\noindent (3) {\it The adjusted perimeter per agent $p$ of the interface between the different agents suitably adjusted for the vacant spaces.} The perimeter $P$ is defined as twice the total number of $R$-$B$ connections plus the total number of connections between $R$ and $B$ agents with vacant spaces, and $p=P/N^2$ (see also discussion in Sect.~2).

\medskip

\noindent (4) {\it The total number of clusters in a configuration $N_C$.} This intuitively appealing measure of aggregation is useful to describe final states having mostly large compact clusters. For such systems, $N_C$ is the quantity that attracts the viewer's attention first. But to immediately see its limitation, observe that ``the maximally integrated'' checkerboard configuration with $v=0$ has just $1+1=2$ clusters. This is because two squares are considered to belong to the same cluster if they touch by a side or a vertex, and clusters may be intermingled. The quantity $N_C$ is the most useful for configurations consisting of compact clusters of a similar size.

\medskip

\begin{figure}[htbp]
\center\begin{tabular}{cc}
\hspace*{-30mm} \includegraphics[width=3in]{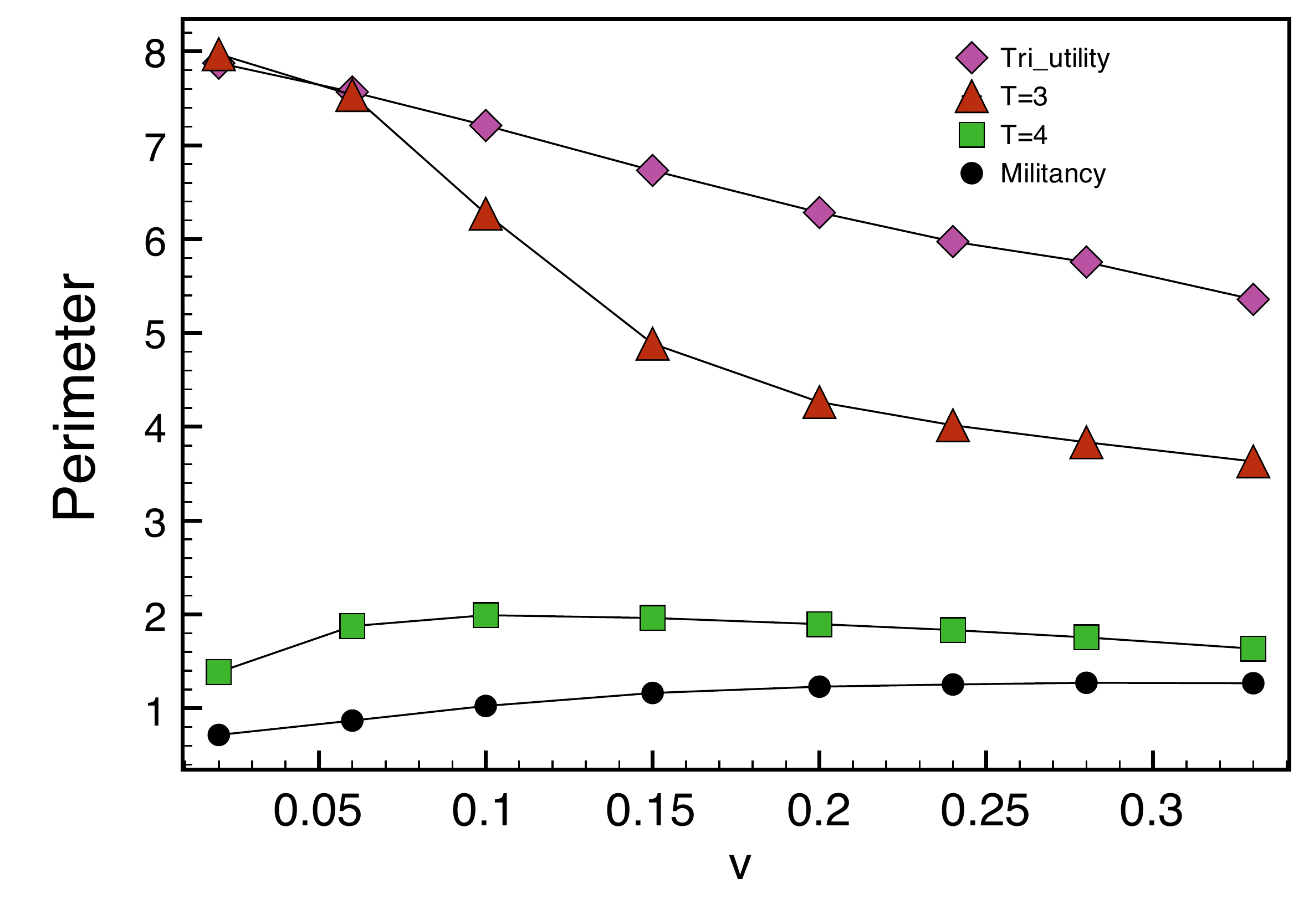}& \includegraphics[width=3in]{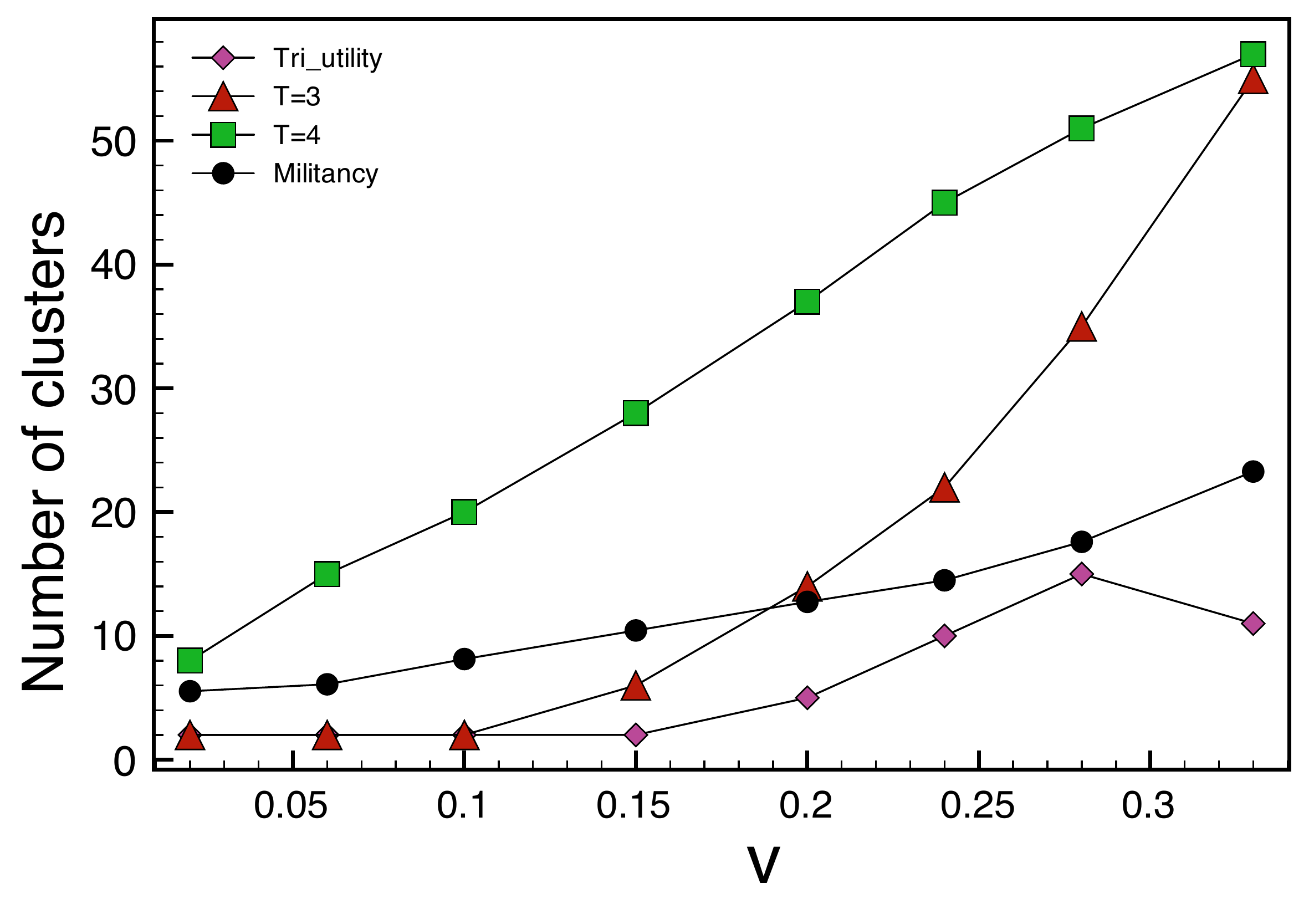}\\
(a)&(b)\\
\hspace*{-30mm} \includegraphics[width=3in]{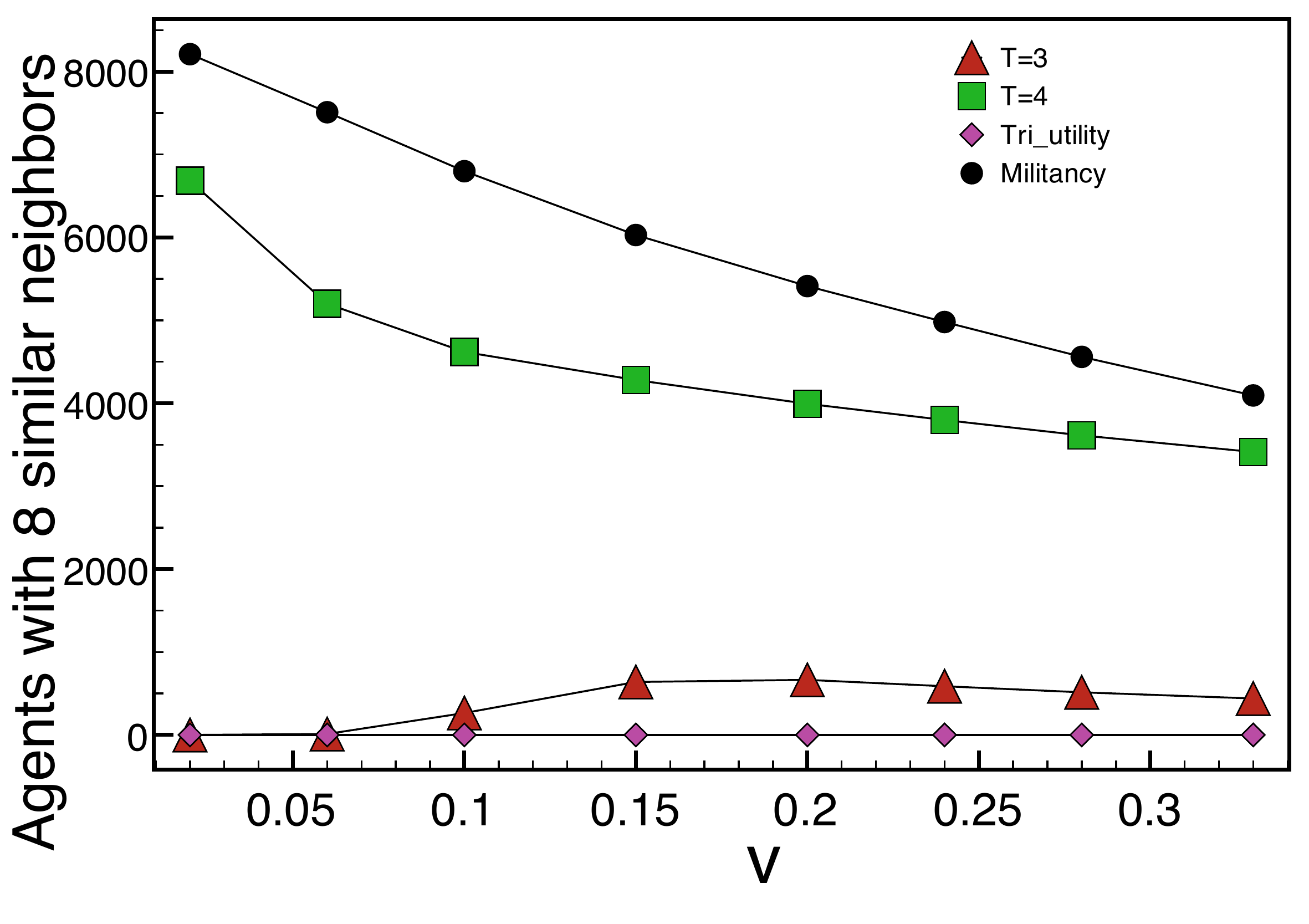}& \includegraphics[width=3in]{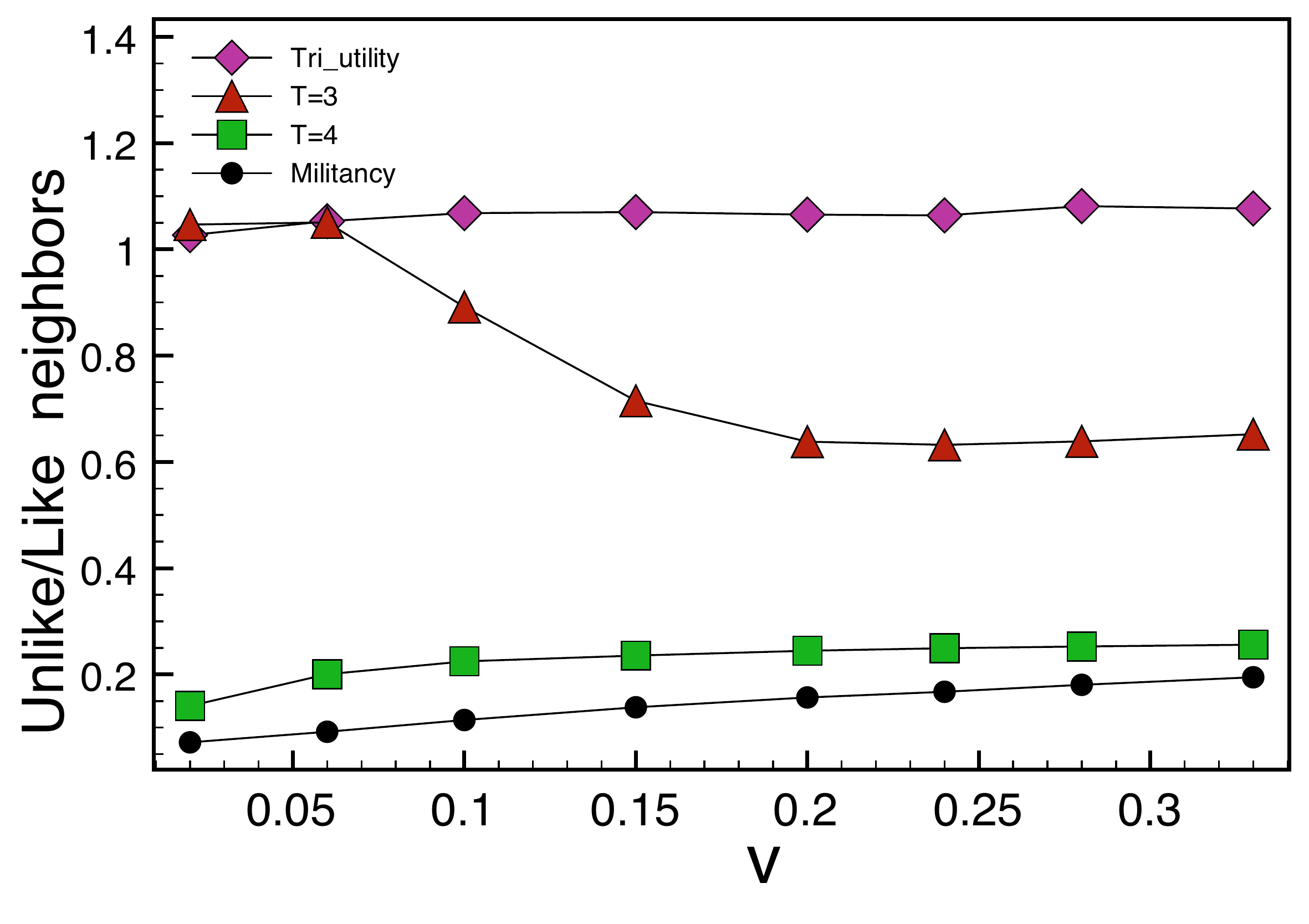}\\
(c)&(d)
\end{tabular}
\caption{\label{Stat} Statistics of final states for Schelling model with regular, militancy and triangle utility functions (a) Perimeter; (b) Number of clusters; (c) Number of agents with $8$ similar neighbors; (d) Unlike/Like Ratio.}
\end{figure}

For each set of parameters' values we run $100$ simulations and Fig.~\ref{Stat} shows plots of average values of these measures of aggregation.


\subsection{Militancy model}

\begin{figure}[htbp]
  \begin{center}
    \includegraphics[width=5in]{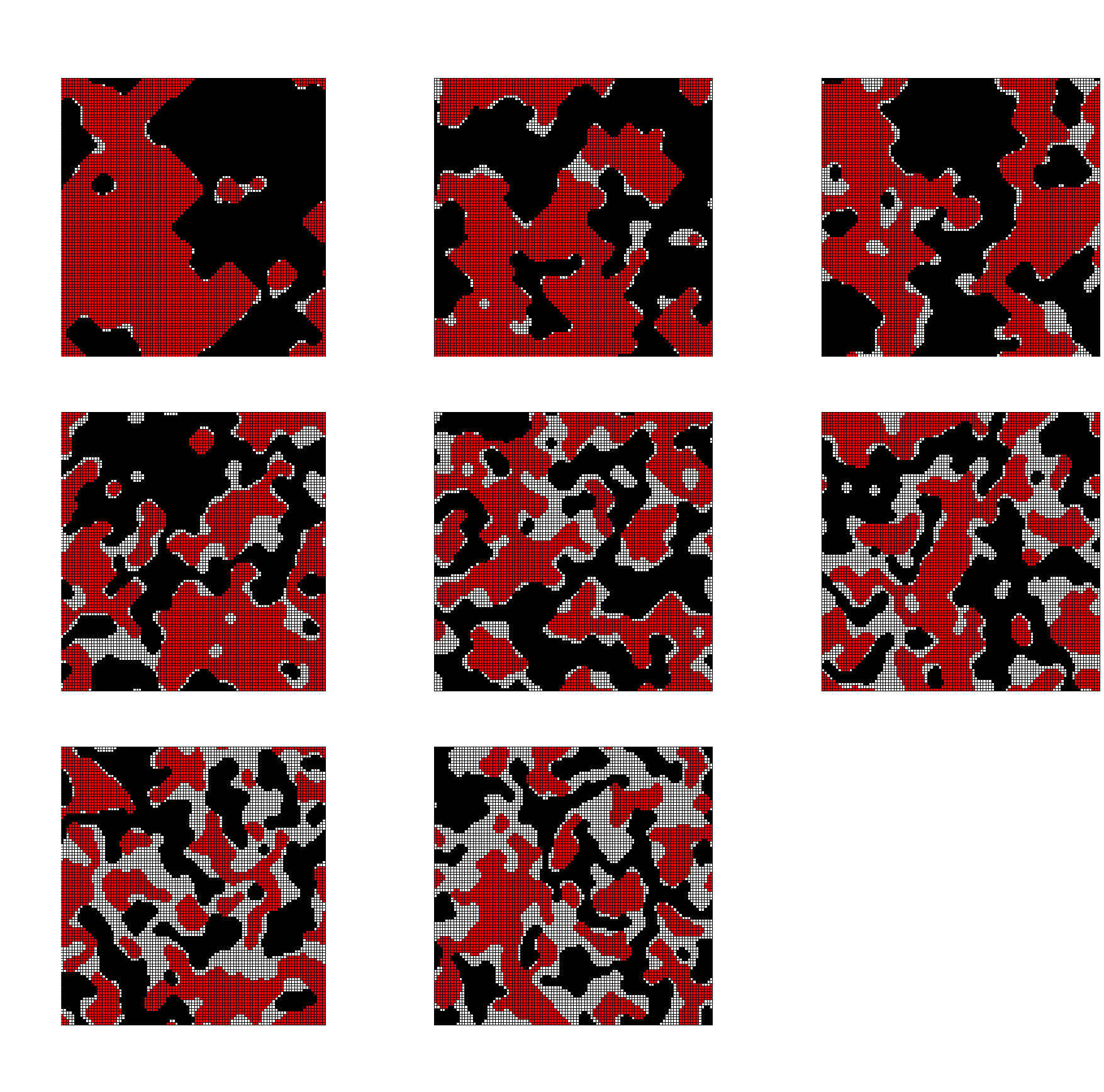}
\caption{\label{mil} Typical final states for militancy model}
  \end{center}
\end{figure}

In some settings individuals may wish to be surrounded by as many neighboring individuals of the same type as possible. Sociologically this could correspond to hostile environments, when the relations between the two types of groups are badly strained -- which is why we called such models {\it militancy models}. Fig.~\ref{mil}. shows characteristic final states for the militancy model.

The weighted total perimeter is a Lyapunov function (see \cite{SVW09} and Sect.~2 for details). However, there is also a simpler Lyapunov function, the sum of the utility functions of all the agents:
\[
L = -\sum U_{i,j}.
\]
Indeed, if a given agent has $s_{i,j}$ like neighbors, the total input in $L$ due to his presence is $-2s_{i,j}$ ($-s_{i,j}$ comes from his personal utility function and $-1$ is contributed by each of his $s_{i,j}$ like neighbors). As every move increases the utility function, $L$ monotonically decreases at every step.

The existence of a Lyapunov function guarantees that the model converges to a final steady state. Moreover, since $L$ decreases by at least one on every switch and $L$ cannot be less than $-16N^2$, there can only be finitely many moves before the algorithm converges to an equilibrium state.

We observe from Fig.~\ref{mil} that for all values of $v$, except for $v=2\%$, the final configuration is far from the global minimum of the Lyapunov function $L$, which is realized when the agents of each kind occupy two completely filled ``strips'' with vacant spots forming a strip between them. The corresponding minimum value of $L_{mim} \apx -16(1-v)N^2 + 12N$. However the Lyapunov function $L$ landscape is filled with local minima and simulations stop when the system reaches any of the local minima. The landscapes for the two Lyapunov functions $L$ and $P$ are not identical. For $L$, every valid switch which decreases $L$ is valid step for the system and vice versa. This is not true for $P$, where although every valid switch decreases $P$, not every move which decreases $P$ is not a valid move. Thus, the reduction of $L$ is the objective of evolution whereas the reduction of $P$ is only an indicator of evolution.


In most simulations, especially for relatively large values of $v$, like agents in final states are contained in one or two large connected clusters that are dense and "snaky" along with at most a few almost circular clusters. The vacant spaces are also "dense" and, frequently serve as buffer zones between the $R$ and $B$ clusters. By providing opportunities for increasingly ``easier satisfaction," one might believe that decreasing $v$ increases the number of $\mbox{centers of aggregation}$. In other words, when there are a lot of vacancies, agents have many choices and it leads to appearance of many small ``islands". Later in evolution, some of the islands may, and do, merge, creating the observed wavy structure. We believe that by allowing a pair of agents, rather than a single agent, to move, the final states may have lower value of $L$.

The statistics of the characteristics of final states for the militancy model resemble those for $T=4$, see Fig.~\ref{Stat}. The difference between them is more quantitative than qualitative. For states with many vacancies, the similarities are the most pronounced, while for small $v$, the final states for the militancy model have much smaller perimeter.

\subsection{Triangle utility function}

\begin{figure}[htbp]
  \begin{center}
    \includegraphics[width=5in]{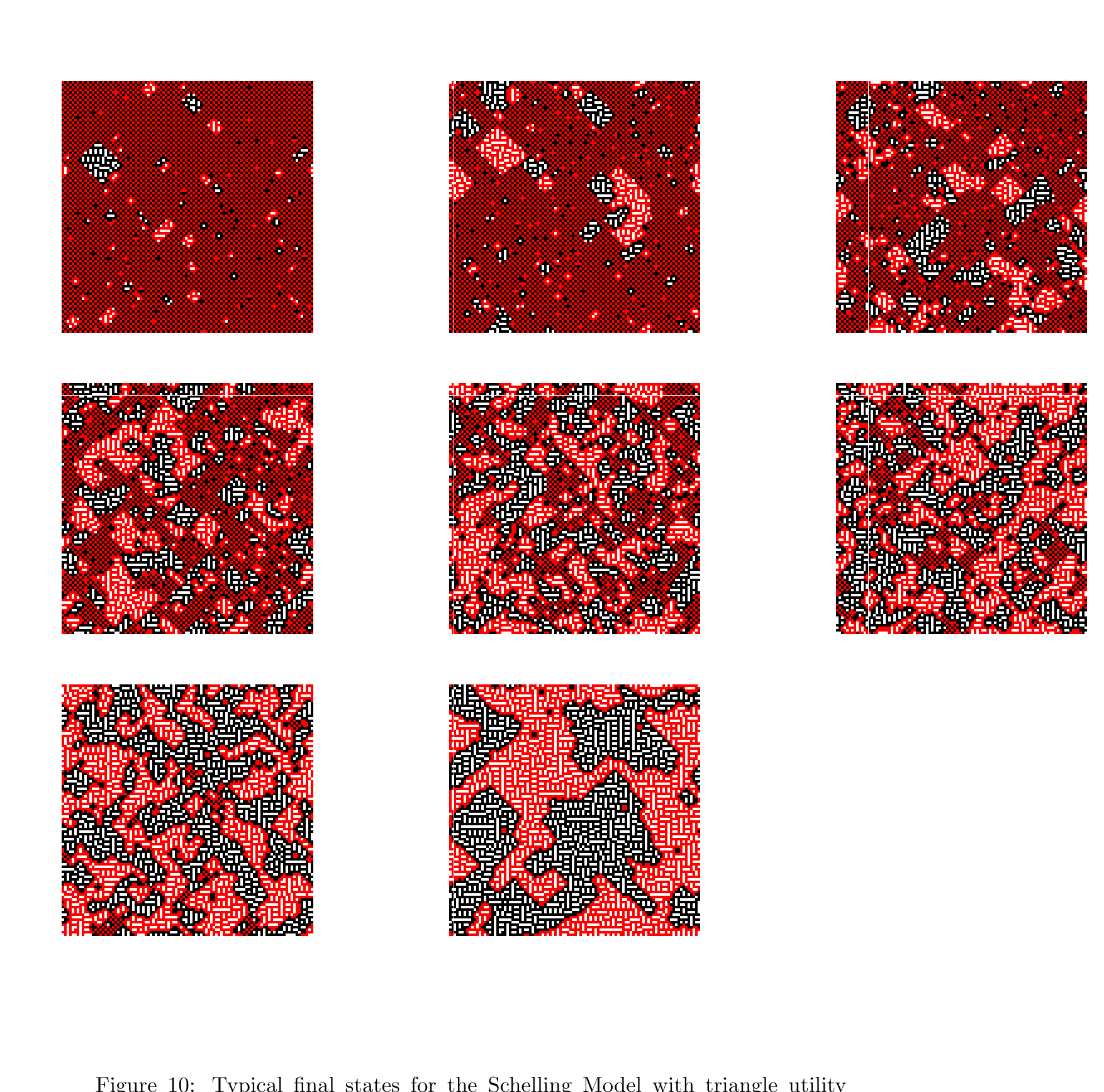}
\caption{\label{finstate} Typical final states for the Schelling Model with triangle utility function for different values of $v$.}
  \end{center}
\end{figure}

Figure~\ref{finstate} illustrates some typical limit states for the triangle utility function.

The clusters in the final states are for most part, intermeshed, but distinct clusters are seen for $v=0.28$ and $v=0.33$. These clusters are not compact and are extremely sparse for $v=0.28$ and $v=0.33$. Therefore the ratio of unlike to like neighbors remains very close to $1$ throughout. Thus the triangle utility allows final states to be less isolated than ones for the threshold models.

Unlike any other case, the final states for the triangular utility function contain clusters possessing a "tessellated-like" structure. The final states are composed of subsets where the original checkerboard configuration survived, islands that contain agents of one type, and vacancies -- all having a type of 'tessellated' structure. As the number of vacancies grow, the 'tessellated' area also grows, reaching the total area around $v=0.28$. As the value of $v$  increases, the islands tend to aggregate into one major cluster of each type. For every value of $v$, the number of clusters is lower than for the regular Schelling Model, which suggests a greater degree of segregation.

The final state statistics resemble those for $T=3$, see Fig.~\ref{Stat}. This is quite natural, since in both cases most agents in the final states have $3$ to $5$ like neighbors.

\end{document}